\documentclass[twocolumn,prb,showpacs,superscriptaddress]{revtex4}
\usepackage{float}
\usepackage{amssymb}
\usepackage{amsmath}
\usepackage{graphicx}
\usepackage{epstopdf}

\usepackage{rotating}
\usepackage{graphicx}
\usepackage{soul}
\usepackage{color}

\begin{document}

%\linenumbers

\def\mean#1{\left< #1 \right>}
%\linenumbers

\title{ \Large The onset, evolution and magnetic braking of vortex lattice instabilities in nanostructured superconducting films}

\author{O.-A. Adami} \affiliation{D\'epartement de Physique, Universit\'e de Li\`ege, B-4000 Sart Tilman, Belgium}

\author{\v Z. L. Jeli\'{c}} \affiliation{D\'epartement de Physique, Universit\'e de Li\`ege, B-4000 Sart Tilman, Belgium}

 \affiliation{Departement Fysica, Universiteit Antwerpen, Groenenborgerlaan 171, B-2020 Antwerpen, Belgium}
 
\author{C. Xue}\affiliation{INPAC -- Institute for Nanoscale Physics and Chemistry, Nanoscale Superconductivity\\ and Magnetism Group, K.U.Leuven, Celestijnenlaan 200D, B--3001 Leuven, Belgium}

\author{M. Abdel-Hafiez} \affiliation{D\'epartement de Physique, Universit\'e de Li\`ege, B-4000 Sart Tilman, Belgium}
 \affiliation{Center for High Pressure Science and Technology Advanced Research, Shanghai 201203, China}

\author{B. Hackens} \affiliation{ NAPS/IMCN, Universit\'{e} catholique de Louvain, B-1348 Louvain-la-Neuve, Belgium}

\author{V. V. Moshchalkov} \affiliation{INPAC -- Institute for Nanoscale Physics and Chemistry, Nanoscale Superconductivity\\ and Magnetism Group, K.U.Leuven, Celestijnenlaan 200D, B--3001 Leuven, Belgium}

\author{M. V. Milo\v sevi\'{c}} \affiliation{Departement Fysica, Universiteit Antwerpen, Groenenborgerlaan 171, B-2020 Antwerpen, Belgium}

\author{J. Van de Vondel} \affiliation{INPAC -- Institute for Nanoscale Physics and Chemistry, Nanoscale Superconductivity\\ and Magnetism Group, K.U.Leuven, Celestijnenlaan 200D, B--3001 Leuven, Belgium}

\author{A. V. Silhanek} \affiliation{D\'epartement de Physique, Universit\'e de Li\`ege, B-4000 Sart Tilman, Belgium}

\date{\today} 
\begin{abstract}

In 1976 Larkin and Ovchinnikov [Sov. Phys. JETP 41, 960 (1976)] predicted that vortex matter in superconductors driven by an electrical current can undergo an abrupt dynamic transition from a flux-flow regime to a more dissipative state at sufficiently high vortex velocities. Typically this transition manifests itself as a large voltage jump at a particular current density, so-called instability current density $J^*$, which is smaller than the depairing current. By tuning the effective pinning strength in Al films, using an artificial periodic pinning array of triangular holes, we show that a unique and well defined instability current density exists if the pinning is strong, whereas a series of multiple voltage transitions appear in the relatively weaker pinning regime. This behavior is consistent with time-dependent Ginzburg-Landau simulations, where the multiple-step transition can be unambiguously attributed to the progressive development of vortex chains and subsequently phase-slip lines. In addition, we explore experimentally the magnetic braking effects, caused by a thick Cu layer deposited on top of the superconductor, on the instabilities and the vortex ratchet effect.

\end{abstract}

\maketitle

\section{Introduction}

Type II superconductors in the mixed state submitted to an external electrical current may exhibit a flow of vortices traversing the superconductor perpendicularly to the current direction.\cite{Kim} In pinning-free samples or at high enough current densities, this flux-flow regime exhibits a linear voltage-current dependence with field-dependent resistivity $\rho(H) \approx \rho_n H/H_{c2}$, where $\rho_n$ is the normal state resistivity and $H_{c2}$ is the upper critical field. When the vortices are driven at high velocities, they may overcome the velocity of sound of the material ($\sim 1$ km/s) giving rise to a \v Cerenkov like hypersound emission.\cite{sound} At even higher velocities ($\sim 10$ km/s), an escape of normal quasiparticles from the vortex core can occur, due to their energy increase above the superconducting gap caused by the electric field. In this case, the resulting quasiparticle depletion in the core reduces the core size and the vortex viscosity. This, in turn, speeds up the vortex motion and leaves behind the vortex a wake of excess quasiparticles. 

In 1976, Larkin and Ovchinnikov (LO) \cite{LO} theoretically predicted that this effect should lead to an instability of the vortex matter and a sudden dynamic transition towards a more dissipative state.
 This prediction has been indeed observed in many different superconducting materials, \cite{Musienko,klein,Doettinger,Samoilov,Doettinger2,Ruck,Xiao,Peroz,Grimaldi,Liang,Grimaldi2,Liang2,Xiao2,silhanekPRL2010,gaia-2015,silhanekNJP,grimaldiAPL} thus demonstrating the universal character of this phenomenon. According to the LO model, the current density $J^*$ and the voltage $V^*$ at the point where the instability is triggered, are uniquely defined. Since within the LO formulation vortex pinning is not considered, the instability takes place simultaneously for all vortices in the lattice, at a  critical vortex velocity $v_c=V^*/BL$, where $L$ is the distance between voltage contacts and $B \sim H$ is the local magnetic flux density. Later refinements of the LO model have permitted to explain the widely observed low field dependent  $v_c \sim 1/\sqrt B$ in low pinning samples \cite{Doettinger} and the switching to an opposity tendency $v_c \sim  H$ when pinning plays an important role.\cite{Grimaldi2,silhanekNJP,grimaldiAPL}

A point that remains unclear in the LO formulation is to what  exactly corresponds the more dissipative dynamic state after the instability has been triggered. This issue has been partially addressed by Vodolazov and Peeters \cite{vodolazov-peeters} using time-dependent Ginzburg-Landau theory {\it in a pinning-free wide superconducting stripe}. It was shown that the LO instability should be preceded by dynamically driven reorganizations of the vortex lattice which manifest themselves as a series of kinks in the current-voltage characteristics $V$($I$). These reorganization transitions are
followed by the development of channels of depleted order parameter populated by kinematic vortices \cite{andronov} or phase-slip lines \cite{sivakov,silhanekPRL2010,1} which coexist with slow-moving Abrikosov vortices. The formation of channels of kinematic vortices is associated with a sudden increase of voltage. Under these circumstances, i.e. in presence of the relatively weak pinning produced by the surface barrier, it is not possible to unambiguously define a unique $J^*$ and $V^*$, as one observes a series of jumps. To date, little is known about whether these theoretically predicted multiple dynamic transitions reflected in  the  $V$($I$) characteristics are still present in realistic samples with pinning centers.

In this work, we study the influence of lithographically defined pinning centers (triangular antidots forming a triangular array) on the vortex instabilities observed at high current densities. This investigation complements our previous work where the pinning strength had been tuned by means of an array of micron size magnetic rings. \cite{silhanekNJP} The main advantage of the present approach with respect to the previous one is that no vortex-antivortex pairs are created due to inhomogeneous fields \cite{Lange, Milosevic, Neal,Kramer} and hence a more straighforward analysis, closer to the theoretical prediction, can be performed. In addition, while in Ref. [\onlinecite{silhanekNJP}] we demonstrated how pinning can give rise to an unexpected field dependence of a well-defined critical vortex velocity, in the present work we unveil experimentally and theoretically the origin of the observed multiple voltage jumps occuring at different critical vortex velocities. In particular, we find that a LO instability as manifested by a unique jump to the normal state at  ($J^*$,$V^*$), can only occur at low fields where the effective pinning is stronger. Varying the magnetic field around the matching conditions where the vortex and hole density coincide, allows us to change the effective vortex pinning and show its influence on the observed dynamic transitions. We conclude that as the effective pinning increases, $V^*$ decreases while $J^*$ remains rather insensitive to pinning.  The final state after the last voltage jump is neither the normal state nor it follows a flux-flow field dependence and it is presumably caused by the strong depletion of the superconducting component of the total current in the region between two neighboring phase slips. This is in agreement with time-dependent Ginzburg-Landau simulations performed for a similar layout as in the experiments. 

Motivated by recent predictions of a damped vortex motion when a metallic layer is deposited on top of the superconductor,\cite{Brisbois} we have also investigated the effects of placing a thick Cu layer on top of the Al bridge but separated by a 50 nm SiO$_2$ film to avoid proximity effects. We observe significant changes in the higher current regime of the voltage-current characteristics due to the Cu layer, but little changes in the low current regime. Interestingly, the presence of the Cu layer also causes a reduction of the ratchet effect which can be attributed to the magnetic coupling between the Al and Cu layers. 

\section{Experimental details}

We investigated two samples, both made of a 50 nm thick Al film deposited by dc-sputtering on a SiO$_2$ substrate. The Al films were patterned with a triangular array of holes (antidots) with a lattice parameter of 3 $\mu$m (see Fig. 1) corresponding to a matching field of 0.276 mT. The shape of the antidots is an equilateral triangle with a side of 0.8 $\mu$m for sample Tr1 and 1.2 $\mu$m for the sample Tr2. The magnetic field was applied perpendicularly to the sample surface ($z$ direction). Transport measurements with the current parallel to the base of the triangles ($x$ direction) were carried out by using a four probe technique as shown in Fig. 1. The width of the transport bridge is 600 $\mu$m and the distance between voltage contacts is $L=$ 2 mm. These large dimensions are needed to guarantee an excellent signal-to-noise ratio. The superconducting critical temperature at zero field was $T_{c0} = $1.364 K for sample Tr1 and 1.374 K for sample Tr2, as determined by a resistance criterion of 0.1$R_n$, with $R_n$ the normal state resistance and using a dc current of 10 $\mu$A. For this applied current, we found a width of the superconducting transition (10\%-90\% criterion) of 5 mK. We estimated a coherence length of $\xi$(0)= 90 $\pm$ 10 nm  and a penetration depth, $\lambda$(0) = 160 $\pm$ 10 nm, showing that these thin Al films are in the type-II superconductivity regime. The nucleation of superconductivity and the ratchet effects in the same structures have been investigated previously.\cite{jorisEPL}

\begin{figure} [h] \centering \includegraphics[width=8cm]{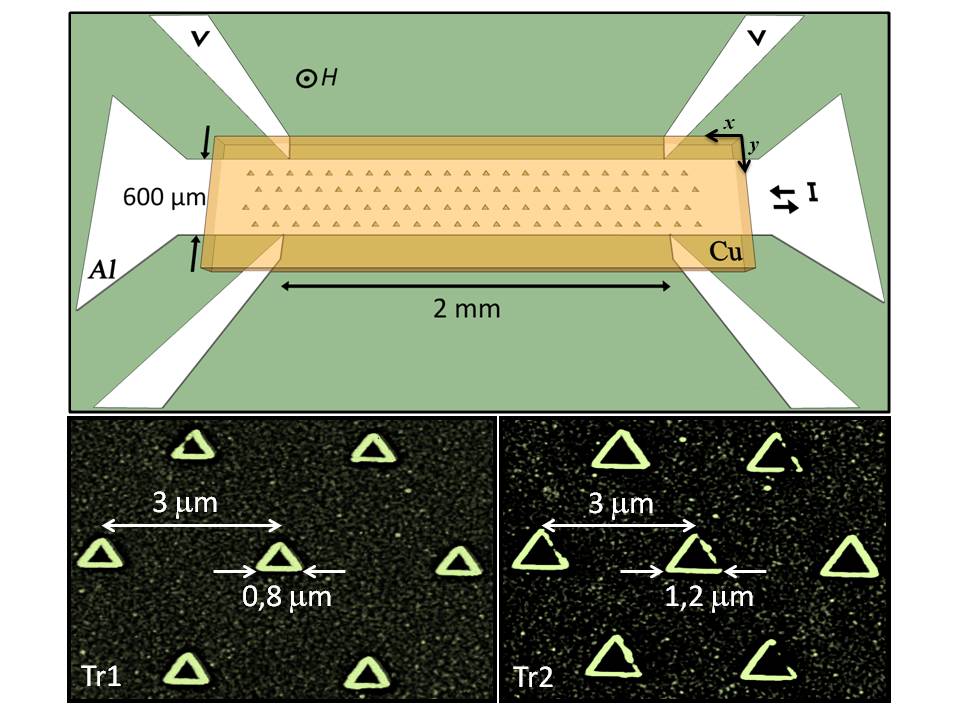} \caption{\label{fig1} Upper panel: sketch of the transport bridge with the corresponding dimensions. Lower panels: scanning electron microscopy images of the Al samples with the triangular array of triangular antidots.} \end{figure}

It is important to mention that the transport measurements have been done with the samples immersed in superfluid $^4$He to minimize heating effects. If heating effects were important, the rate at which the $V$($I$) curves are collected should play an important role. The fact that for these samples the $V$($I$) characteristics are always the same, irrespective of whether the measurements are performed with pulsed currents ($\mu$s) or by slow sweeps (ms), confirms that heating effects can be neglected. In addition, the low current regime (i.e. the continuous part of the $V$($I$) characteristics) is perfectly reproduced when sweeping up and down the current. In view of the generality of the results obtained for $T>0.96T_c$, we present in this work data corresponding to a single temperature $T=0.97T_c$. Since the superconducting parameters are functions of the reduced variable $T/T_c$, working on both samples at the same reduced temperature ensures a more reliable comparison between them. 

A sister sample, Tr2b, was covered with 50 nm SiO$_2$ layer deposited by plasma-etched chemical vapor deposition to produce a conformal coating of the patterned Al film. This is a crucial step needed to electrically isolate the Al layer from a subsequent 500 nm thick Cu layer deposited on top by dc-magnetron sputtering (see Fig. \ref{fig1}). The perfect insulation between the two layers (Al and Cu) is consistent with (i) the absence of proximity effect as evidenced by the fact that no significant difference in the critical temperature is observed with and without Cu layer, and (ii) no change in the normal state resistance of the Al layer, and therefore no electrical shunt observed. Furthermore, we also measured on the Al film after removing the Cu layer and verified that the same properties as the original Al film were reproduced.

\section{Theoretical modeling}
In the numerical simulations we considered a thin Al film (effectively two-dimensional) of size $L_1 \times L_2 = 13~\mu $m$ \times 24~\mu $m,  in the presence of a perpendicular magnetic field $H$ and with a transport current density $J$ applied through  normal-metal leads. The sample has been patterned with a triangular array of the triangular holes with size $a=$1.2 $\mu$m, and pinning period of 3 $\mu$m so as to match with the experimental sample Tr2. Numerical data have been analysed for a single temperature $T=0.97T_c$. We considered a coherence length at zero temperature $\xi_{Al}(0) =$ 90 nm, from where it follows that the coherence length at the working temperature $\xi_{Al}(T)=\xi_{Al}(0)/\sqrt{1-\frac{T}{T_c}}$= 520 nm.

To explore the dynamic properties of the superconducting condensate in this system, the time-dependent Ginzburg-Landau (tdGL) equation has been solved in its generalized form:\cite{watts,1}
\begin{equation}
\label{eq1}
\begin{split}
\frac{u}{\sqrt{1+\gamma^2|\Psi|^2}}\left(\frac{\partial}{\partial t}+i\varphi+\frac{\gamma^2}{2}\frac{\partial |\Psi|^2}{\partial t}\right)\Psi= \\
(\nabla-i\textbf{A})^2\Psi+(1-|\Psi|^2)\Psi,
\end{split}
\end{equation}
coupled with the equation for electrostatic potential:
\begin{equation}
\label{eq2}
\Delta\varphi=\nabla\left(Im\left\lbrace\Psi^*(\nabla-i\textbf{A})\Psi\right\rbrace\right).
\end{equation}
Both equations are given in dimensionless form, where the order parameter $\Psi$ is given in the units of $\Delta(T)=4k_BT_cu^{1/2}/\pi\sqrt{1-T/T_c}$ ($k_B$ being the Boltzmann constant, and $u=5.79$ the ratio of the relaxation times of the order parameter amplitude and phase). The distances are expressed in the units of coherence length $\xi$, time is scaled with Ginzburg-Landau relaxation time, $\tau_{GL}=\pi\hbar/8uk_B(T_c-T)$. The magnetic field is scaled by $H_{c2}=\Phi_0/2\pi\xi^2$, where $\Phi_0$ is the flux quantum. In these units, the electrostatic potential $\varphi$ is given in units of $\varphi_0=\hbar/2e\tau_{GL}$, the magnetic vector potential is scaled by $\xi H_{c2}$, and the current is given in units of $j_0=\sigma_n\hbar/2e\tau_{GL}\xi$ ($\sigma_n$ is the normal-state conductivity). The parameter $\gamma=10$ is a product of inelastic collision time for electron-phonon scattering and $\Delta(T)$. 
At the superconductor-vacuum boundaries, the following conditions have been used: $(\nabla-i\textbf{A})\Psi|_n=0$ (zero supercurrent), and $\nabla\varphi|_n=0$ (zero normal current). At the boundaries where the normal metal leads are assigned, the boundary conditions are $\Psi=0$ and $\nabla\varphi=-\textbf{J}$. 

All simulations were performed in the same manner. For each value of magnetic field $H$, we started the simulations multiple times from randomly generated Cooper-pair distribution. By utilizing the least free energy criterion we determined the ground state from the random states.\cite{golib} The obtained ground states then served as an initial solution to which the current is subsequently introduced and gradually increased, until the sample transits to the normal phase. From these simulations the $V$ ($I$) characteristics were obtained. Voltage is measured between first and fourth fifths of the sample width $L_2$ as indicated in Fig. Fig. \ref{tdGLVI}.The total duration of the simulation for each set of the parameters was $t_{sim}=7000\tau_{GL}$, which typically provided enough time for the system to reach a dynamic equilibrium.

\section{Results and discussion} 

\subsection{ Multi-step $V$($I$) characteristics} 
In order to identify the different vortex dynamic regimes we measured $V$($I$) characteristics at several applied fields and temperatures. In Fig. 2(a) a set of $V$($I$) curves taken at several magnetic fields is shown for the sample  Tr1. Similar behavior has been found for the sample Tr2.
Fig. 2(b) shows a magnification in the low voltage regime of the data presented in Fig. 2(a). As current densities increased, several distinctive dynamic phases can be identified. The critical current $J_c$ separates the pinned vortex phase from the flux-flow regime and is determined by using a 1 $\mu$V voltage criterion, or equivalently, a $0.5$ V/m electric field criterion. In the flux-flow regime and irrespectively of the density of vortices, the $V$($I$) curves do not follow a purely linear dependence mainly due to a non-zero width distribution of vortex velocities and to a less extent due to a current dependent effective pinning energy.\cite{comment0} It is worth noting that within this regime of the  $V$($I$) characteristics, the very same curve is reversibly followed if the current sweep is stopped at a certain current and the sweep is reversed.   

\begin{figure} [h] \centering \includegraphics[width=8cm]{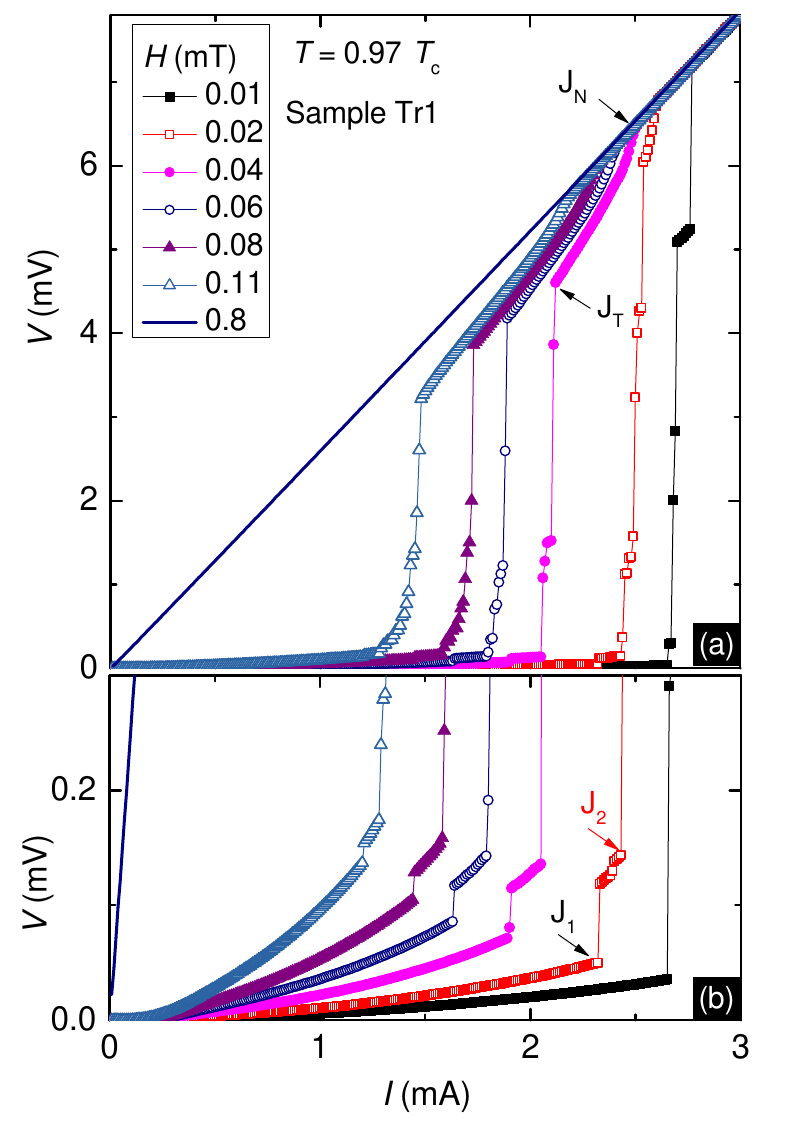} \caption{\label{fig2} (a) Experimental current-voltage characteristics for sample Tr1 obtained at $T=0.97T_c$ and for several applied magnetic fields. The magnetic field $H=0.8$ mT is above the upper critical field and the curve corresponds to the Ohmic response in the normal state. (b) Zoom-in of the low voltage regime. The onset of different dynamic phases is indicated by arrows.} \end{figure}

The continuous increase of voltage in the flux-flow regime is suddenly interrupted at a current density $J_1$ by an abrupt jump towards a more dissipative regime. Within this new dynamic phase the voltage monotonously increases as the drive increases until a new instability current density $J_2$. In between $J_1$ and $J_2$, the absolute resistance $R=V/I$ is not constant but increases with current. This first plateau can be followed by subsequent jumps of irregular sizes until $J=J_T$ when the system enters in a peculiar highly dissipative regime. The fact that the differential resistance $dV/dI$ in this high current regime exhibits a weak magnetic field dependence, contrasting with the typical flux-flow behavior, suggests that a different dissipative mechanism than vortex motion should be considered. Eventually, the sample smoothly transits to the normal state at $J=J_N$ where all curves collapse. As the applied magnetic field approaches the upper critical field, the $V$($I$) characteristics become smoother and the instability points are no longer well defined. Decreasing the current from a value $J>J_1$, will lead to hysteretic irreversibility with a retrapping current substantially smaller than $J_1$. 

\begin{figure} [h]  \includegraphics[width=9.5cm]{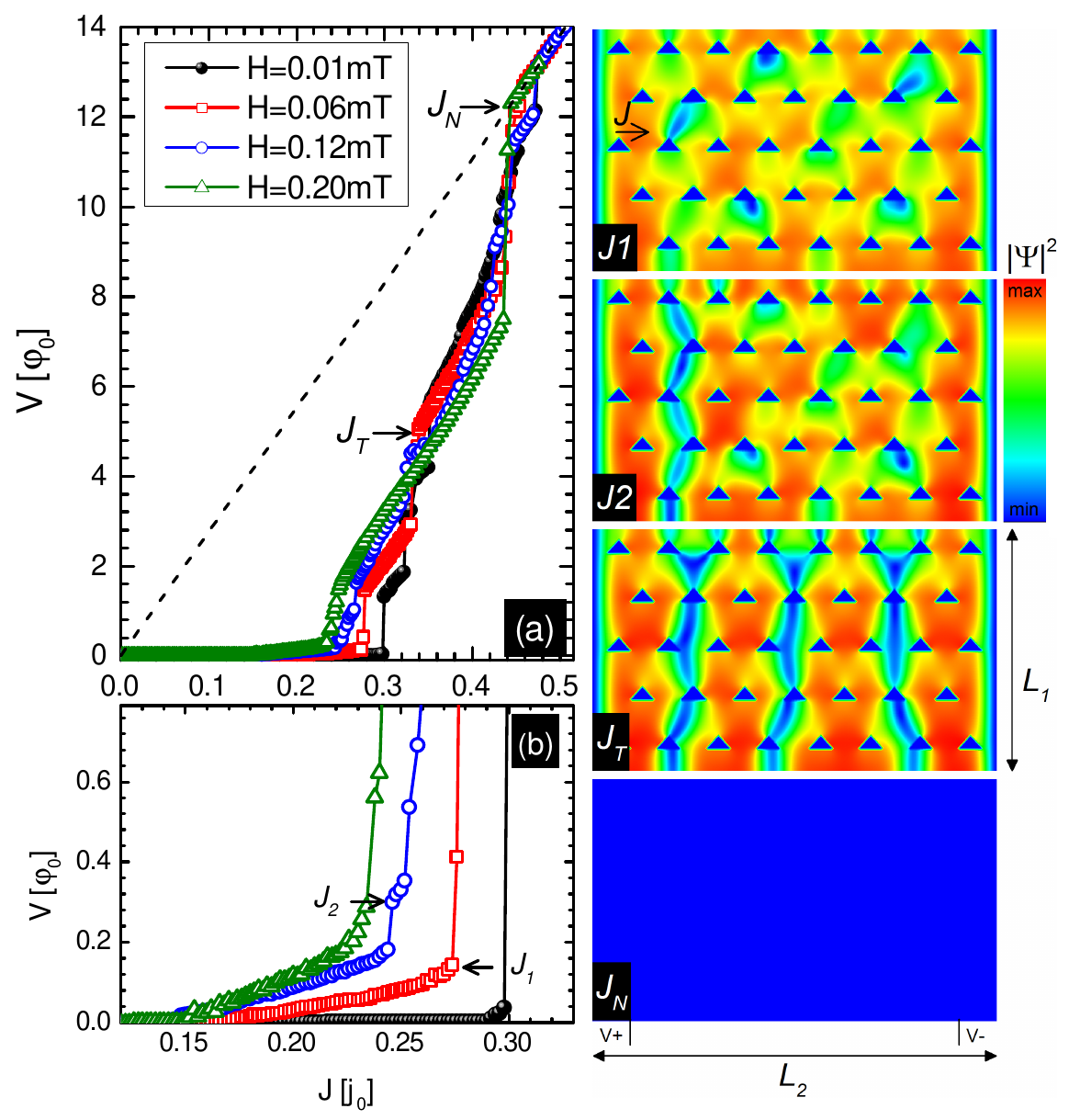} \caption{\label{tdGLVI} (a) Simulated current-voltage characteristics for sample Tr2 at $T=0.97T_c$ and for several applied magnetic fields. (b) Zoom-in of the low voltage regime. The onset of different dynamic phases is indicated by arrows. The right column shows four contour plots of the Cooper-pair density for different values of the applied current flowing from left to right at a magnetic field $H \approx 0.2 H_{c2}$. In these images, vortices cross the bridge from top to bottom. In the lower right panel the contacts between  which the voltage is measured are indicated. } \end{figure}

A summary of the $V(I)$ characteristics along with snapshots of the superconducting Cooper-pair density obtained by using time-dependent Ginzburg-Landau numerical simulations is presented in Fig.   \ref{tdGLVI}. For computational convenience, the simulated system is substantially smaller than the experimental samples. Nevertheless, most of the dynamical regimes identified in the experimental data, delimited by $J_1$, $J_2$, $J_T$, and $J_N$, still have a clear counterpart in the modeled electrical response.

  Simulations also unveil the ultimate origin of the observed voltage jumps. Namely, the first  dissipative voltage jump at $J_1$ takes place when the first path of depleted order parameter percolates from one side to the other of the transport bridge. Just before the first jump (panel $J_1$), the vortex cores exhibit clear deformation, becoming elongated in the direction of motion. These deformed vortices represent the so-called kinematic vortices.\cite{andronov,1} In the current interval $J_1<J<J_2$, a well defined channel of depleted order parameter (blue stripe in panel $J_2$) is formed by the continuous passing of kinematic vortices. Note that along this channel the order parameter is not fully suppressed, as evidenced by the non uniformity in the color intensity. The observed jump in the $V(I)$ characteristic at $J_2$  is a consequence of the development of another stripe of kinematic vortices, appearing adjacent to the previously existing one. Depending on the distance between voltage contacts, several subsequent channels can be formed as current increases, leading to a series of voltage jumps. Eventually, with the repeated passing of faster kinematic vortices along the channels, the order parameter cannot heal anymore and a smooth transition to what is known as phase-slip line (PSL) takes place.\cite{sivakov,1} This regime of PSL corresponds to the current range above $J_T$ and below $J_N$.
  
 It is well documented that phase slips avoid each other, the reason being that at the PSL most of the applied current $J=J_n+J_s$, transforms into normal current $J_n$, and since vortices are only driven by the supercurrent $J_s$, those sitting close to a PSL undergo a slower motion.\cite{vodolazov-peeters} This may also be the reason for the long resistance plateau observed for  $J_T<J<J_N$. In this current range, a maximum possible number of PSLs have been established, and the sample can be regarded as a series of channels of fully depleted order parameter, separated by channels of restored order parameter. The latter will exist until the critical supervelocity  $v_s=J_N/(2e|\psi|^2$)  is reached, typically at current $J_{N}\leq J_{dp}$. It is worth stressing the importance of an efficient heat removal in order to achieve this coexistance of channels with depleted order parameter separated by superconducting regions. 

\subsection{ Field-dependence of dynamic phases }
\begin{figure} [h!] 
\centering 
\includegraphics[width=8.5cm]{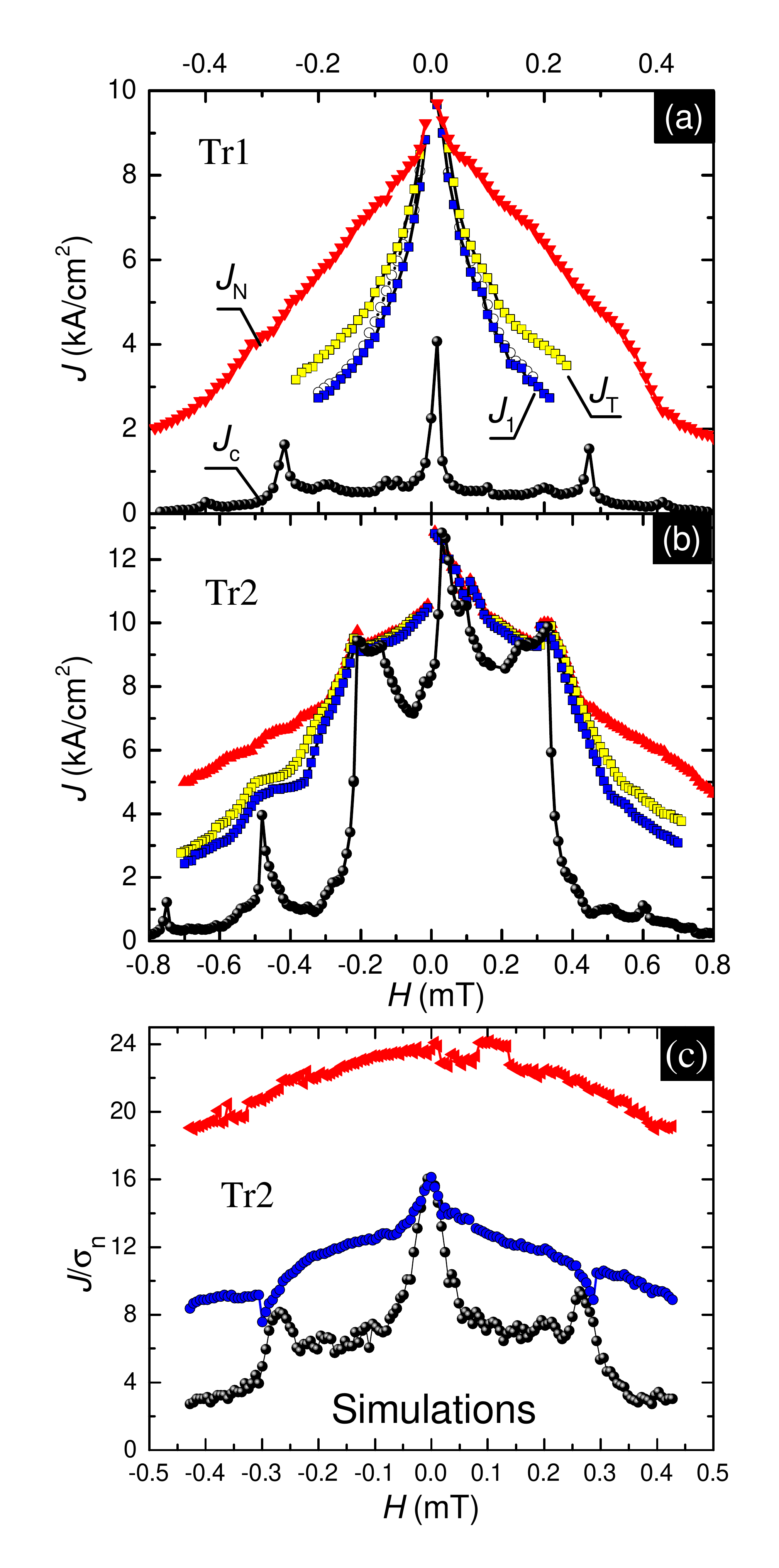}
\caption{\label{fig4}Dynamic vortex phases as a function of applied current ($I$) and magnetic field ($H$) for samples (a) Tr1
(b) Tr2 experimental, and (c) Tr2 simulated. $J_{c}$ is the depinning current, $J_{1}$ and $J_{T}$ indicate the first and last voltage jumps, respectively, and $J_{N}$ is the current needed to reach the normal state. In panel (a), the open circles indicate the end point of the first plateau $J_2$. } \end{figure}

Let us now discuss the field dependence of the different phases introduced in the previous section. In Fig. \ref{fig4}(a-b) we show the experimentally determined current-field dynamic phase diagram for both samples Tr1 and Tr2 at $T=0.97T_c$. The black dots mark the collapse of the vortex pinning phase at $J_c(H)$, which exhibits local maxima at zero field and at every field where the vortex lattices commensurate with the pinning landscape. Note that the sample with larger holes (Tr2) gives rise to a higher critical current density, consistent with the fact that pinning strength increases with hole size.\cite{buzdin,nordborg-vinokur,moshchalkov, Berdiyorov, Berdiyorov1}
 For the sample Tr1, at fields higher than the first matching field, i.e. when the vortices outnumber pinning sites, a rather continuous transition to the normal state is observed and an unequivocal determination of the voltage jumps is no longer possible. In addition, the $J=J_N$ transition is insensitive to the critical current enhancements at the matching fields, thus suggesting that this line is unlikely to be related to vortex pinning. This is in agreement with the proposed interpretation of this transition as linked to the deparing current. This is further confirmed by the dynamic phase diagram obtained by the simulations and presented in Fig. \ref{fig4}(c).

For sample Tr2 at high fields, $J_N$($H$) exhibits a similar behavior as for Tr1, i.e. $J_N$($H$) is insensitive to the matching fields for $|H|>0.4$ mT. However, for $|H|<0.4$ mT, there is a single voltage jump in the $V$($I$) characteristics, directly to the normal state, all transitions collapse in one line, and matching features become also imprinted in the $J_N$($H$). In Tr2 when $|H|<0.4$ mT the flux-flow regime is also absent at the first matching field. In other words, the multiple steps transitions turn into a single step transition at the matching field. This finding indicates that single voltage jumps can be recovered in the strong pinning limit, however, in contrast to the LO model, this single jump is not preceded by a free flux-flow but rather a highly non-linear $V$($I$) characteristic. It is also worth noting that the asymmetric shape of the holes leads to a slightly different dynamic response for opposite field polarities, hence pointing out the importance of vortex pinning on the high-velocity vortex instability.

The dynamic phase diagram obtained by the simulations and presented in  Fig. \ref{fig4}(c) reproduces most of the features shown by the experimenatl data. In particular, it confirms that the current of the first voltage jump,  $J_1$, is not very sensitive to the effective pinning variations.  Nevertheless, considering that the inequality $J_1>J_c$ must be always satisfied, the sudden increase of the effective pinning at the matching fields will inevitably be reflected in the $J_1$ curve. This is in agreement with the experimental data presented in Fig. \ref{fig4}(a-b) and with earlier reports based on magnetic pinning landscapes.\cite{silhanekNJP} We stress that in both experiment and simulations that $J_1 = J_c$ at matching fields. In other words, flux flow is completely suppressed by pinning, and the system transits directly to the LO state. 

\subsection{Pinning-dependent dissipation upon instability}

\begin{figure} [h]
\centering
\includegraphics[width=8cm]{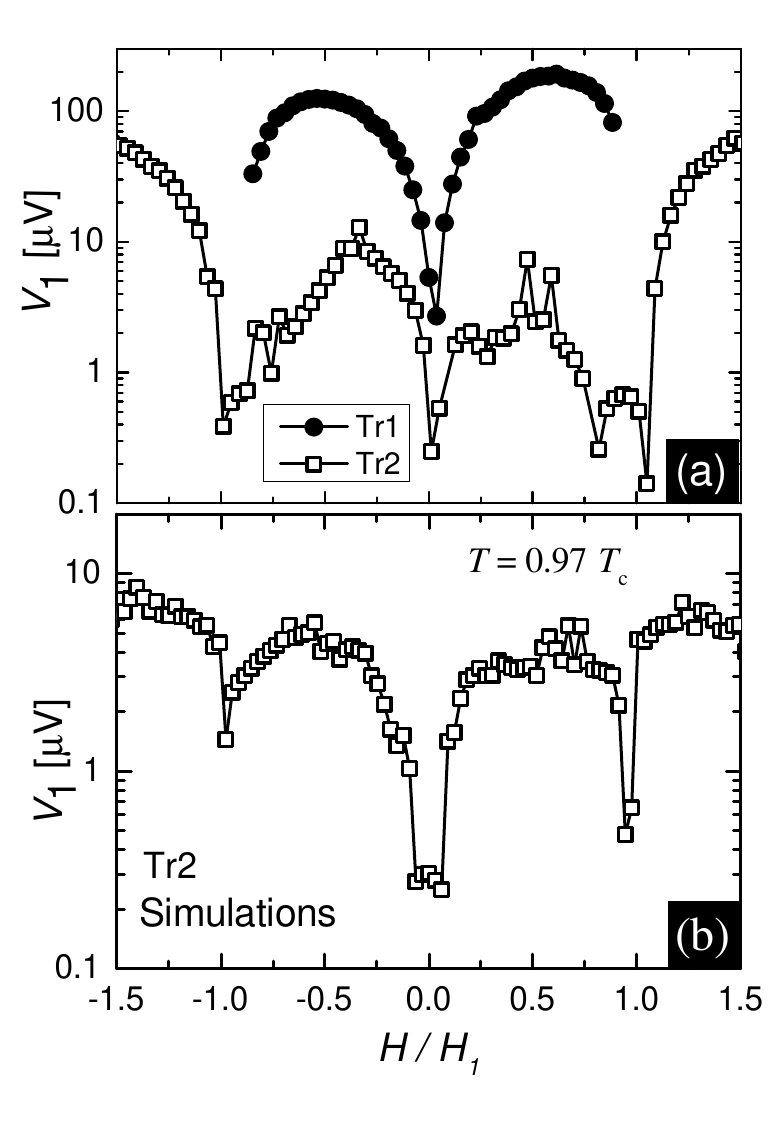}
\caption{\label{fig5} Voltage of the first instability point $V_1$ as a function of the magnetic field for the two samples Tr1 and Tr2
at $T=0.97T_c$ (a), and for the simulated system (b).} \end{figure}

In addition to the instability current density $J_1$, it is also interesting to track the evolution of the instability voltage $V_1$. Assuming that all vortices move with equal velocity $\bf v$, the electric field induced by the vortex motion is $\bf E = B \times v$, and therefore knowing the instability voltage $V_1=|\bf E \cdot L|$, it is possible to obtain the critical velocity $v^*=V_1/BL$, at which the instability is triggered. Larkin and Ovchinnikov proposed an explicit expression for $v^*$($T$) from where the quasiparticle inelastic scattering time can be extracted.\cite{Doettinger,doettinger-tsuei} The LO theory assumes a homogeneous distribution of quasiparticles from where a field dependence of the critical velocity, $v^*$($B$) $\propto$ $1/\sqrt{B}$ was predicted.\cite{Doettinger} The latter expression works well at high enough fields but fails to reproduce the increase of the instability voltage with increasing field, at low $B$, i.e. when vortex pinning is important. We have recently argued that this discrepancy results from the fact that pinning, not considered in the original LO theory, leads to highly non-linear $V$($I$) characteristics which in turn invalidates the use of the relation $v=V/BL$ to extract the critical velocity. Indeed, every time vortex pinning becomes important, the vortex lattice instability takes place within a plastic regime characterized by a highly inhomogeneous vortex velocity distribution with a reduced fraction of moving vortices, $n$($B$). As a consequence, the instability voltage $V_1 \propto n$($B$)$v^*$ should also be reduced. This interpretation is in agreement with the field dependence of the voltage $V_1$ shown in Fig. \ref{fig5}(a) where, for matching conditions, i.e. when the effective pinning is enhanced, a dip in the critical voltage $V_1$ is observed. These results reinforce and extend the findings reported in Ref. \cite{silhanekNJP} to the case of non-magnetic pinning and are further confirmed by the tdGL results presented in Fig. \ref{fig5}(b).

\subsection{Magnetic braking realized by Cu coating}

It is already well known that a metal of low electrical resistivity and high thermal conductivity in contact with a superconductor can improve the thermal and dynamic stability of the superconductor.\cite{wilson} This is the reason why most of the commercially available superconducting cables consist of superconducting filaments (typically NbTi, Nb$_3$Sn, or V$_3$Ga) immersed in a conducting matrix (Cu or Al). This hybrid superconductor-metal structure is also present in the new generation of high-$T_c$ cables.\cite{bernstein,larbalestier-natmat} As early as in 1974, Harrison {\it et al.} \cite{Harrison0, Harrison1} reported  a reduction of the size and the speed of the magnetic flux jumps in the superconductor due to an adjacent metallic layer. The effect was not only attributed to the thermal conductivity of the metallic layer (or heat sinking effect) but also to the magnetic coupling between the flux motion in a superconducting film and the eddy currents induced in the metallic layer. Later on, with the development of the magneto-optical imaging technique and the simulation tools,  similar effects were observed in different superconducting thin films with a metallic capping layer.\cite{Baziljevich,Stahl, JVestgarden,Choi0,Choi1, Colauto,Albrecht, Brisbois} Thus far, however, little is known concerning the influence of such electromagnetic braking on the LO instability. A first attempt to tackle this issue was carried out by Peroz {\it et al.} \cite{Peroz1} who studied the evolution of the LO instability in a superconducting film in presence of proximity effect due to a metallic layer. They observed a reduction of the critical velocity and the critical force at which the flux-flow instability appears. This effect could be attributed to the influence of the metallic layer on the relaxation time of the nonequilibrium quasiparticles which governs the LO mechanism. Unfortunately, the fact that the coherence length $\xi$, the effective penetration depth $\Lambda = 2 \lambda^{2}/d$, the critical temperature $T_{c}$ and  the second critical field $H_{c2}$  are also affected by the proximity effect makes it impossible to isolate the contribution of the electromagnetic coupling. Interestingly, Danckwerts {\it et al.}\cite{danckwerts} showed  in Pb films that a single voltage jump-up (probably due to LO instability) is not influenced by a 2D conductive layer, whereas the retrapping current (i.e. jump-down voltage) is shifted to higher current values. 

\begin{figure} \centering \includegraphics[width=9cm]{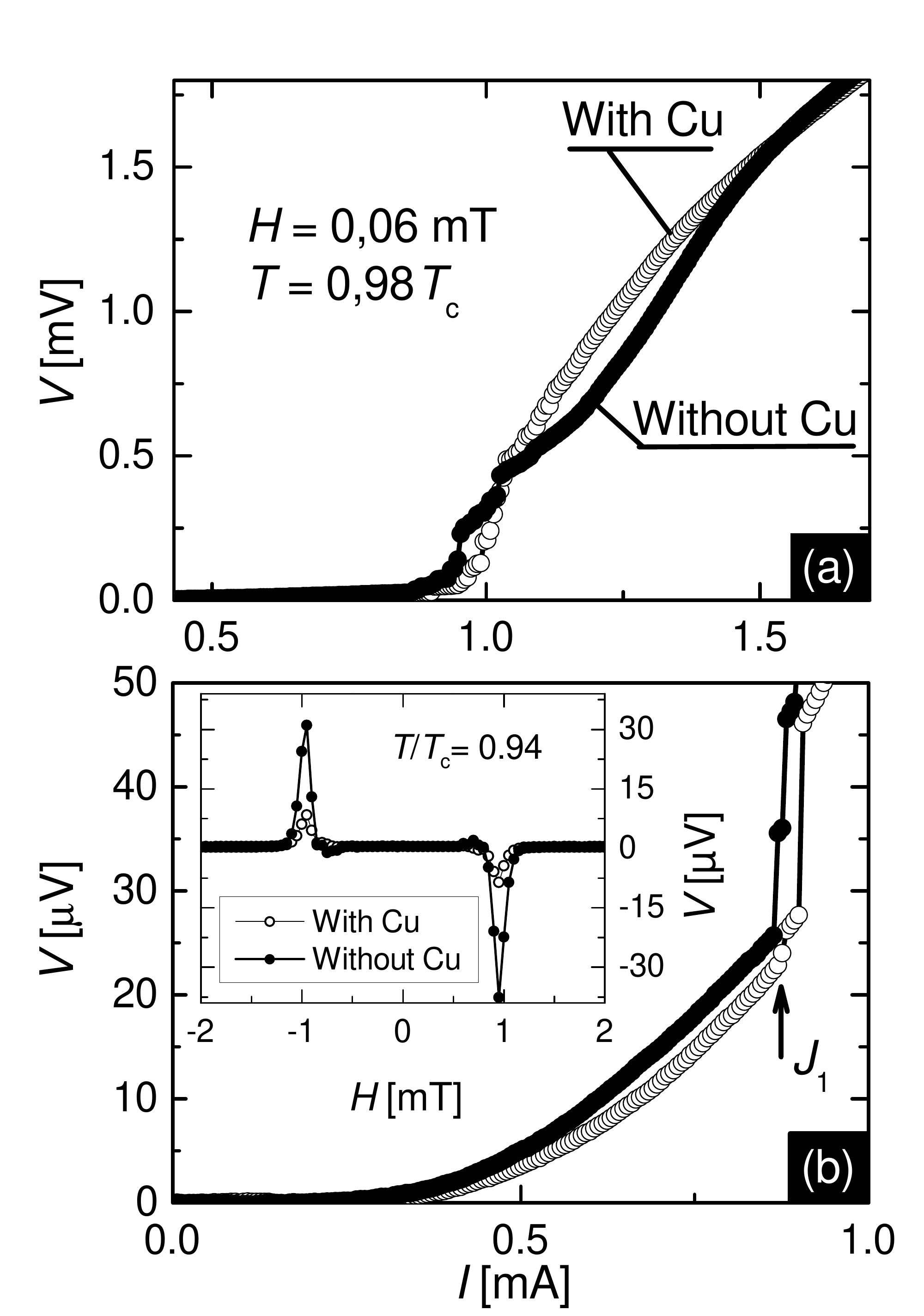} \caption{\label{fig6}  (a) Voltage versus current for the sample Tr2 with and without a 500 nm thick Cu layer on top. Panel (b) shows a low voltage zoom-in of the panel (a) evidencing the same instability point $J_1$ with and without Cu. The inset in panel (b) shows the rectified voltage V$_{dc}$ versus magnetic field when applying an alternative current of 1 mA and frequency of 33 kHz at $T/T_{c}$ = 0.94.} \end{figure}

\begin{figure} \centering \includegraphics[width=8cm]{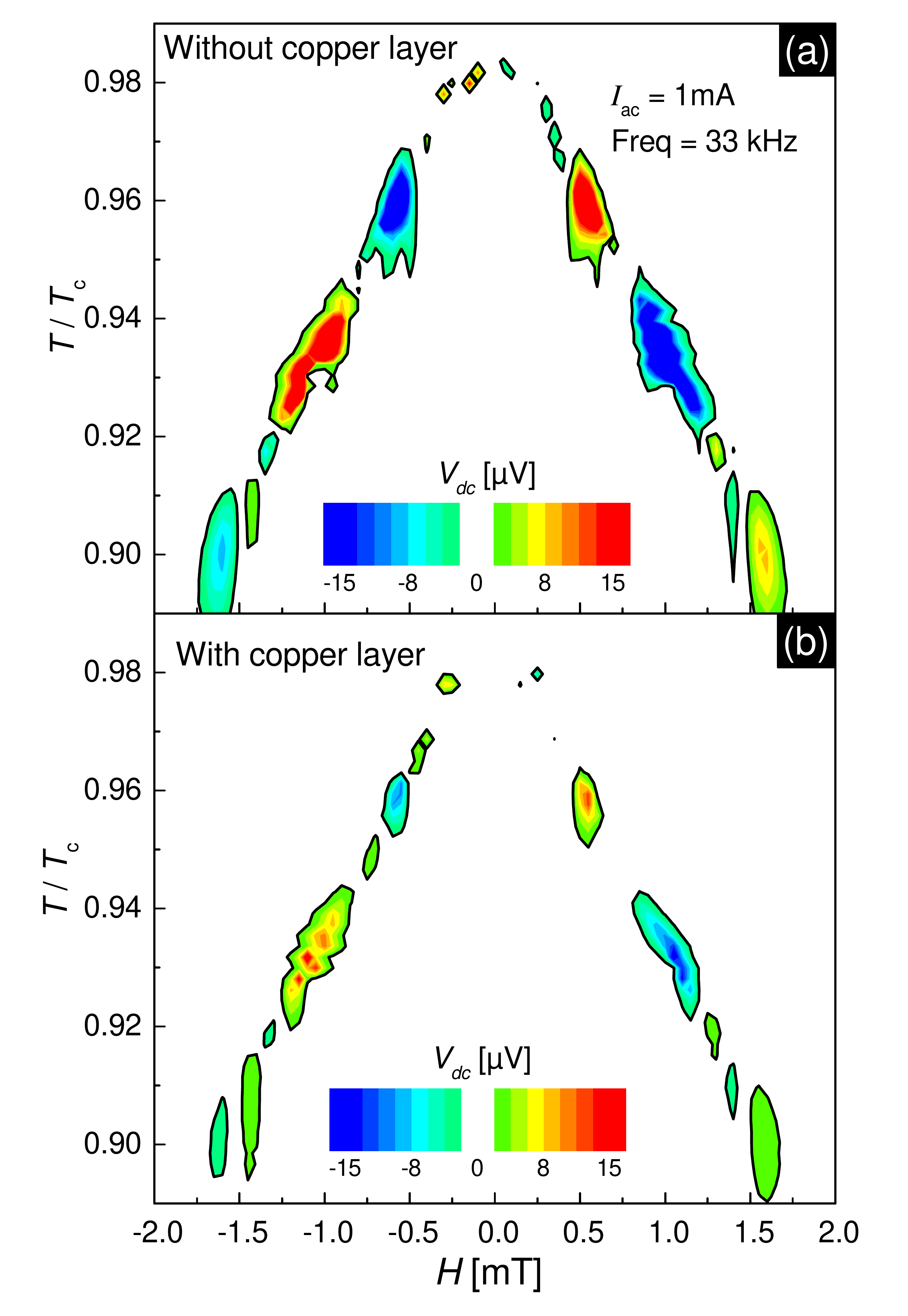} \caption{\label{fig7}  Phase diagram of the vortex ratchet effect. Contour plot of the magnetic field and temperature dependence of the measured dc voltage V$_{dc}$ with an ac current of 1 mA and frequency 33 kHz applied to sample (a) without Cu layer and (b) covered with a Cu layer. } \end{figure}

In this section, we provide further insights on the influence on the LO instability as well as the vortex ratchet motion, of an electrically insulated (i.e. without proximity effect) Cu layer on top of the Tr2b sample (see experimental section).  Comparing the $V(I)$ characteristics of the Tr2b sample with and without a 500 nm thick Cu layer on top (Fig.\ref{fig6}), we found that the first instability point at $J_1$ remains nearly unaffected by the presence of the copper layer as already reported by Danckwerts {\it et al.}\cite{danckwerts} This is shown in Fig. \ref{fig6}(b) for $H=0.06$ mT at $T=0.98T_c$ while similar behavior is observed for other temperatures and fields. We note that the dynamic phase developping at higher current densities than $J_T$ for the Al film without Cu, and previously attributed to the channels of depleted order parameter, is substantially modified when the sample is covered with Cu (see  Fig. \ref{fig6}(b)). The absence of this regime for the Cu coated sample is most probably due to a more efficient spreading of the heat (initially produced along the vortex channels) into the superconducting layer, thus accelerating the transition to the normal state. 

An alternative way to investigate the influence of the Cu layer on the vortex dynamics consists of measuring the rectification of the vortex motion produced by the asymmetric pinning sites. Indeed, when an alternating current $J=J_{ac}\sin(2\pi ft)$ is applied, a net vortex flow develops due to the fact that the critical current becomes polarity dependent, i.e. $J_c^x \neq J_c^{-x}$. In other words, an $ac$ excitation of zero mean value will give rise to a finite $dc$ voltage $V_{dc}$. Assuming $J_c^x < J_c^{-x}$, if $J_{ac}<J_c^x$, then $V_{dc}=0$, i.e. the shaking current is not large enough to depin the vortex. When $J_c^x<J_{ac}<J_c^{-x}$, the considered vortex can hop from site to site during one half period, and remains immobile during the other half period of the oscillation. In this case,  $V_{dc} \neq 0$. For even larger current amplitudes such that $J_{ac}>J_c^{-x}$ the vortex can move in both directions, but still completing a longer trajectory in one particular direction. The excess current acting on the vortex leads to a vortex velocity given by, $\eta v = (J-J_c) \Phi_0 $, where $\eta$ is the damping coefficient and $v$ is the instantaneous vortex velocity.\cite{comment1} From this relation, assuming that $J_c$ is not influenced by the Cu layer (as expected) and that the voltage $V_{dc}$ is proportional to the average vortex velocity, $v$, it is easy to show that,

\begin{equation}
\label{damping}
\frac {V_{Al}}{V_{Al+Cu}} = \frac {\eta_{Cu} + \eta_{Al}}{\eta_{Al}},
\end{equation}  

\noindent where $V_{Al}$ is the average $dc$ voltage measured in the Al film, without Cu on it, and $V_{Al+Cu}$ corresponds to the $dc$ voltage measured in the Al film with a Cu layer on top. The term $\eta_{Al}$ is the vortex damping in the superconductor as described by the Bardeen-Stephen model,\cite{BS} and $\eta_{Cu}$ represents the additional damping produced by the magnetic braking induced by the eddy currents in the Cu layer.\cite{danckwerts,Brisbois}

Eq. (\ref{damping}) tells us that the ratchet signal ($V_{dc}$) should be smaller for the sample with Cu layer, simply because in this case vortices move in a medium with higher viscosity. This is indeed in agreement with the experimental data reported in Fig. \ref{fig7}, where the $V_{dc}$ is plotted as a function of magnetic field and temperature for a sinusoidal excitation of amplitude $I=1$ mA and frequency $f=33$ kHz. Fig. \ref{fig7}(a) shows the results for the bare Al film, Fig.\ref{fig7}(b) corresponds to the Al film covered by the Cu layer, and in the inset of Fig. \ref{fig6} a comparison of both ratchet signals taken at constant reduced temperature is presented. 
 
Taking the maximum rectification voltage for 50 different temperatures in the range $0.88 < T/T_c < 0.98$, we estimate the ratio $V_{Al}/V_{Al+Cu} \approx 2.6 \pm 1.6$, which implies $\eta_{Cu}/ \eta_{Al} \approx 1.6$. Within this temperature range no significant change of $V_{Al}/V_{Al+Cu}$ is observed. The damping coefficient\cite{danckwerts} $\eta \propto \sigma d/a^2$, where $\sigma$ is the material conductivity, whereas $a$ and $d$ are the characteristic length scales in-plane and out-of-plane, respectively,  where normal dissipative currents circulate. For the superconducting layer, Bardeen-Stephen model\cite{BS} suggests that $a \approx \xi(T)$ and $d=d_{Al}=50$ nm is the total thickness of the Al film. For the Cu layer, eddy currents circulate in a region of radius given by the field variation scale length, so for very low fields (i.e. individual vortices) $a \sim \Lambda(T)$ but for moderate or high fields this distance is set by the vortex spacing $a \sim \sqrt{\Phi_0/B} \ll \Lambda(T)$. The measured ratchet signal peaks at magnetic fields where this inequality is always satisfied. Assuming that the mean vortex speed is low enough such that the electromagnteic skin depth is larger than the Cu thickness $d_{Cu}=500$ nm,\cite{saslow} we obtain  $\sigma_{Cu}d_{Cu}B\xi(T)^2/ \sigma_{Al}d_{Al}\Phi_0 \approx 1.6$. Notice that the ($H$,$T$) points of maximum rectification signal $V_{dc}$ follow a nearly linear dependence in temperature, implying that $B \propto 1-t$ which renders the factor $B\xi(T)^2$ temperature independent. Taking the extremes of the investigated phase space where the ratchet signal maximizes, ($H$,$t$)=(1.6 mT, 0.90) and ($H$,$t$)=(0.5 mT, 0.98), we obtain $\sigma_{Cu}/ \sigma_{Al}\sim 2.8$.

Let us now evaluate if the assumption of low mean vortex speed taken above is reasonable. The skin depth\cite{saslow} of the magnetic field into the Cu layer is given by $\delta=\sqrt{wdh/v}$, where $w \sim 100$ m/s is the speed of decaying eddy currents, $d=$500 nm is the Cu thickness,\cite{Brisbois} $h$ is separation between the Cu layer and the superconductor, and  $v$ is the average vortex velocity during the ratchet motion with Cu layer on top. The mean vortex speed can be deduced from the dc voltage $V_{dc}=HLv$. From the inset of Fig. \ref{fig6}(b) for the sample with the Cu on top, we observe $V_{dc} \approx 10 \mu$V for $H = 1$ mT, which yields $v \sim 5$ m/s. This mean vortex velocity implies a skin depth $\delta \sim 700$ nm which is indeed larger than the thickness of the Cu, in agreement with our assumption.

\section{Conclusion}

In conclusion, we have shown that in realistic superconductors with ever present pinning, voltage jumps appear as the current density increases as a consequence of the progressive developement of channels of depleted order parameter likely populated by kinematic vortices. This phase slowly evolves into a dense array of phase-slip lines across the sample. Within this scenario, it is no longer possible to unambiguously refer to a unique Larkin-Ovchinnikov vortex instability at high vortex velocities. A single voltage jump is recovered if the pinning strength is strong, however, unlike the LO model, in the weaker pinning case the instability is triggered in a highly non-linear voltage-current regime. Interestingly, an enhancement of the vortex viscosity can be achieved by covering the superconductor with a thick layer of normal metal. This effect has also been demonstrated by the substantial reduction of vortex ratchet at large $ac$ excitations.

Although in the present work a particular symmetry of the pinning lattice and individual pinning sites has been chosen, the overall response of the system, the dynamic phases, and the main conclusion of the manuscript should be independent of these details. Nevertheless, the actual current range of each dynamic phase and the morphology of the vortex channels, will be influenced by the symmetry of the pinning array. If, for instance, a square array was used instead of the triangular one, the paths of depleted order parameter, as shown by the snapshots of Fig. 3, would follow the direction of the principal axes of the array. This particular case, with one of the principal axes coinciding with the Lorentz force, has been discussed in Ref. \onlinecite{silhanekPRL2010}. Concerning the symmetry of the individual pinning motifs, by using a triangular symmetry we are able to demonstrate that, as we stated above, the details of the dynamic phases are influenced by the particular geometry of the holes. In particular, it is possible to find a regime where quasi phase-slip lines are triggered for one current polarity, but they are absent for the opposite polarity. A similar effect has been reported in Ref. \onlinecite{vandevondel2011}.

\section{Acknowledgements} This work was partially supported by the Fonds de la Recherche Scientifique - FNRS, the Methusalem Funding of the Flemish Government, the Research Foundation-Flanders (FWO), and COST Action MP1201. The work of A.V.S. and Z.L.J. is partially supported by ``Mandat d'Impulsion Scientifique" MIS F.4527.13 of the F.R.S.-FNRS. The authors thank Jo Cuppens for the data analysis at the early stage of this work, R. Delamare for his valuable help during the fabrication of the samples, and G. Grimaldi for helpful discussions.
%\\


\begin{thebibliography}{99}
\bibitem{comment0} This is so since vortex creep caused by thermal excitation is typically weak in low $T_c$ superconductors, and particularly in Al.
\bibitem{comment1} This equation of motion for a vortex does not take into account the possible deformation of the vortex core that developes at high currents.
\end{thebibliography}
\end{document}